# Optical probing of pups' brain tissue and molecular specific nuclear nano-structural alterations due to fetal alcoholism via dual spectroscopic approach

Prakash Adhikari,[1] Pradeep K. Shukla,[2,†] Shiva Bhandari,[3] Avtar S. Meena,[2] Binod Regmi,[1] Fatemah Alharthi,[1] Peeyush Sahay,[3] Radhakrishna Rao,[2] and Prabhakar Pradhan [1,*]

[1]Department of Physics and Astronomy, Mississippi State University, Mississippi State, MS, USA, 39762
[2] Department of Physiology, University of Tennessee Health Science Center, Memphis, TN, USA, 38103
[3] Department of Physics, East Carolina University, Greenville, NC, USA, 27858

*pp838@msstate.edu
† pshukla2@uthsc.edu

**Abstract:**

Mesoscopic physics-based dual spectroscopic imaging techniques, partial wave spectroscopy (PWS) and inverse participation ratio (IPR), are used to quantify the nano to submicron scales structural alterations in postnatal pups' brain cells/tissues due to fetal alcoholism. Chronic alcoholism during pregnancy, being teratogenic, results in fetal alcohol syndrome and neurological disorder. Results of PWS studies of brain tissues show a higher degree of structural alterations. Furthermore, the IPR analyses of cell nuclei show that spatial molecular mass density structural disorder increases in DNA while decreases for histone. This study characterize the brain spatial structures from molecular to tissue level in fetal alcoholism.

## 1. Introduction

Light is an important probe used to detect the structural properties of cells/tissues. The nano to submicron scales intracellular structural alterations take place inside the cells, mainly in macromolecules such as DNA, RNA, lipids etc. due to chronic alcoholism, disease, or other abnormalities. These structural changes in macromolecules of cells result in structural change in tissues. A mesoscopic physics-based technique, the partial wave spectroscopy (PWS) that has a sensitivity of nanoscale changes can distinguish mass density fluctuations in terms of refractive index fluctuations due to cancer, stress or alcohol in cells/tissues [1–4]. Besides, the structural changes in tissues, there are different levels of molecular specific structural alterations that happen in the cell nuclei are related to disease or abnormalities. The molecular specific photonic localization technique, IPR is a sensitive technique recently used to probe and quantify the structural alterations in cells due to cancer, alcohol or other drugs [5–8], including alterations in the brain cells.

Alcohol consumption has been a part of human culture since the beginning of human civilization, however, chronic alcoholism is responsible for several health issues in the present world. Many major health problems reported are directly or indirectly concerned with alcohol. The damage from alcohol, in cells and tissues of a particular organ, depends on various factors like age, sex, amount, and concentration of alcohol, etc. Alcohol affects all the major organs of the body like a liver, heart, brain, pancreas, etc. The immune system is degraded by chronic alcohol consumption and the body is prone to disease thereafter [9]. Therefore, alcohol consumption is a major cause of increased nausea resulting in death over a long period of time [10]. Researchers have shown that a complex and multidimensional relationship occurs between alcohol consumption and health consequences [11]. Chronic alcoholism is, therefore, an important factor responsible for different health problems. The Center for Disease Control and Prevention estimates a death toll of 88,000 annually from alcohol related causes, and alcohol is ranked as the third leading preventable cause of death in the US. The number of deaths due to alcohol is increasing year by year.

Alcohol has been identified as a carcinogen and is responsible for a majority of cancers [12]. Most of the research in alcohol has shown that it has more effects on women than on men [13]. Chronic alcoholism of a mother during pregnancy, or fetal alcoholism, is one of the major risk factors of mental retardation in the children born to such mothers in the United States and worldwide [14]. Fetal alcoholism leads to several critical health issues like miscarriage, stillbirth, and different disabilities known as fetal alcohol spectrum disorders (FASDs) [15]. Fetal alcohol syndrome (FAS) and alcohol-related neurodevelopmental disorder (ARND) are serious outcomes of fetal alcoholism and linked with notable cognitive and behavioral deficits to a newborn child [16]. The one-child born from every 13 exposed to fetal alcoholic mother found to have fetal alcohol disorder syndrome (FADS). The number of infants born with FADS is estimated to be 630,000 annually worldwide and this number is increasing every year [17]. This issue, which is easily preventable, is tragic as it is a leading cause of intellectual disability, birth defects, and developmental disorders [18]. This demands for early and accurate quantification of structural abnormalities due to fetal alcoholism in pup brains' cells and tissues.

Several factors are responsible for the complication in analyzing the effects of alcohol exposure on fetal brain development. The condition changing the blood alcohol concentration (BAC) in the fetus plays an important role in influencing the occurrence and harshness of alcohol-induced developmental brain-injuries [19]. The alcohol consumed by the pregnant mother passes through the placenta to the growing baby in the womb and consumes the same amount of alcohol as the mother. The baby is unable to metabolize the alcohol through its liver or any other organs and alcohol, being teratogenic, interferes with the healthy growth of vital organs resulting in brain damage and other birth defects. These brain cells/tissues are affected at the nano to submicron levels which look similar to the normal can't be predicted by the ongoing histological procedures. The structural alterations in brain tissues and cell nuclei are not well studied or understood.

We report the study of the mesoscopic physics-based dual spectroscopic imaging techniques, the PWS and Confocal-IPR, which can provide a plethora of information of the structural changes at submicron scale levels in tissues and nuclei of brains of mice pup exposed to fetal alcoholism. First, we probe the nanoscale refractive index fluctuations in thin sections of brain tissues using the PWS technique, that quantify the degree of structural disorder at the tissue level. Furthermore, we examine the spatial structural changes in the molecular spatial mass density of nuclear components, DNA(DAPI stained) and histone protein (stained histone protein), using the IPR technique via confocal imaging. The details of PWS study of tissue samples and IPR study of DNA and histones of cell nuclei and obtained results, and their significance and applications are described below.

## 2. Method

### 2.1 Brain cell and tissue samples preparation

All animal experiments were performed according to the protocol approved by the University of Tennessee Health Science Center (UTHSC) Institutional Animal Care and Use Committee (IACUC). Mice were housed in groups of 2-5 per cage, segregated by sex, in a room on a 12 h/12 h light/dark cycle (lights on at 8:00 AM, off at 8:00 PM) maintained at 22±2 °C. Pregnant mice for 8 weeks were fed with ethanol (0% 2 days, 1% 2 days, 2% 2 days, 4% 1 week, and 5% 1 week) in Lieber-DeCarli liquid diet (EF). The control group was pair fed with an isocaloric diet. Brains from control or pair fed (PF) and ethanol fed (EF) offspring were collected at 60-day postnatal age and cryofixed for fluorescence staining using confocal microscopy. While brain tissues from the offspring of pair fed (PF) and ethanol fed (EF) mice were excised cryofixed and sectioned to a thickness of 10µm using microtome for the PWS and IPR analyses.

### 2.2 Partial wave spectroscopy (PWS)

#### 2.2.1 PWS instrumentation

Partial wave spectroscopy is a recently introduced highly sensitive structural disorder imaging technique that combines mesoscopic physics-based analysis with spectroscopic images to detect the nanoscale to submicron scales structural alterations in cells/tissues. A detailed experimental explanation of the PWS instrumentation is presented in previous work [3,4,20,21]. In brief, a stable broadband white light source, Xenon Lamp (150W) is passed through the 4f combination of lenses and gets collimated. Using a right-angle prism, the collimated beam is reflected towards the samples through the objective lens (40X, Newport). A highly sensitive XYZ motorized stage (X-Y axis 40nm and Z-axis 100nm, Zebar

Technologies) is used to focus the collimated beam into the sample within the working distance of the objective. The resulting visible spectroscopic backscattered signal is projected toward the CCD camera filtered with the liquid crystal tunable filter (LCTF, Kurios) which has a resolution of 1nm. Finally, for every wavelength in the visible range (450-700nm) of backscattered (reflected) signals are recorded in the CCD camera. The CCD camera and LCTF are triggered by the LCTF controller to record the signal in the CCD at each wavelength.

*2.2.2 Calculation of disorder strength, $L_d$*

In the PWS technique, the sample is virtually divided into a collection of parallel channels each with diffraction-limited transverse size, to detect the backscattered waves propagating along quasi 1D trajectories within these channels. The backscattered signals at every wavelength *(λ)* and at each spatial pixel position *(x,y)* are recorded in the CCD camera as *R(x,y,λ)* and then the statistical properties of the nanoarchitecture of tissue are quantified by analyzing the fluctuating part of the recorded intensities. The collected backscattered signals are applied with a Butterworth filter and suitable order polynomial to remove the noise. Then, the degree of structural disorder strength, $L_d$, is calculated from the rms value of the extracted intensities $<R(k)>_{rms}$ and the wave vector auto-correlation decay of the reflection intensity ratio, *C(Δk)*. At the spatial pixel position *(x,y)*, the degree of structural disorder strength, $L_d$ is given as [1,3,20,22]:

$$L_d = \frac{Bn_0^2 \langle R \rangle_{rms}}{2k^2} \frac{(\Delta k)^2}{-\ln(c(\Delta k))}\Big|_{\Delta k \to 0} \qquad (1)$$

Where B is the normalization constant, $n_0$ is the average refractive index of the cells/tissues, and $k$ is the wavenumber ($k = 2\pi/\lambda$). This disorder strength can be further simplified and expressed as the product of the variance and the spatial correlation length of the refractive index fluctuations, $L_{d(PWS)}=<\Delta n^2>l_c$.

*2.3 Confocal microscopy for DNA and histone spatial molecular mass density structures*

*2.3.1 DNA and histone staining in cell nuclei in tissue samples*

The immunofluorescence staining was performed in cell nuclei of brain sections with two types of molecular mass densities: DNA and histone.

1) *DNA staining:* Nuclear DNA treated with DAPI, a nuclear dye that binds to mainly DNA, in turn, chromatin. Tissue sections are washed 3 times for five minutes in PBS and mounted on a glass slide using Prolong Diamond antifade mountant containing DAPI. The DAPI present in the mountant is a DNA binding dye that enables to visualize nuclei by fluorescence microscopy.

2) *Histone staining:* Immunofluorescence staining of histone was performed using the H3K27me3 antibody. Cryosections (10 µm thick) were fixed in one to one mixture of acetone and methanol at -20°C for 2 minutes and rehydrated in phosphate buffered saline (PBS). PBS is a mixture of 137 mM sodium chloride, 2.7 mM potassium chloride, 10 mM disodium hydrogen phosphate and 1.8 mM potassium dihydrogen phosphate. Sections were permeabilized with 0.2% Triton X-100 in PBS for 10 minutes and blocked in 4% non-fat milk in Triton–Tris buffer (150 mM sodium chloride containing 10% Tween-20 and 20 mM Tris, pH 7.4). It was then incubated for 1 hour with the primary antibodies (rabbit polyclonal anti-H3k27me3), followed by incubation for 1 hour with secondary antibodies (cy3-conjugated anti-rabbit IgG antibodies).

*2.3.2 Confocal imaging*

Confocal imaging was performed on DAPI and H3K27me3 antibody stained nuclei of cells in their native states in tissues, collected from the frontal cortex region of a mouse model. The fluorescence was examined using a Zeiss710 confocal microscope (Carl Zeiss Microscopy, Jena, Germany), and images from x–y sections (1m) were collected using LSM 5 Pascal software (Carl Zeiss Microscopy). Images were stacked by using the software ImageJ (NIH, Bethesda, MD), and processed by Adobe Photoshop (Adobe Systems Inc., San Jose, CA). All images for DNA and histone tissue samples from different groups were collected and processed under identical conditions. The images obtained were categorized into four groups as (i) PF-DAPI stained brain cells, (ii) EF-DAPI stained brain cells, (iii) PF-H3K27me3 stained brain cells, and (iv) EF-H3K27me3 stained brain cells.

*2.3.3 Inverse participation ratio (IPR) and molecular specific structural disorder analysis*

The method of confocal imaging and quantification of light localization properties of the biological system has been described previously and the details are given elsewhere [5,23]. The fluorescence emitted by the sample molecules from small finite volume or voxel *dV (i.e. dxdydz)* around the excitation center is collected by the photodetector. The confocal image intensity *I(r)* is found [23] to be $I(r) \propto dV(\rho)$ where *I(r)* is the pixel intensity of the confocal image at position *r*, *ρ* is the density of the molecules at small volume *dV*. It has been shown that the local refractive index of a cell is proportional to its local mass density [23–25], i.e. *n(r) =n$_o$ + Δn(r)*. So, a representative refractive index matrix represented by the fluorescence molecules mass density variation is constructed using the pixel intensity values, with the optical potential *ℰ(x, y)* defined as:

$$\varepsilon(r) = \frac{\Delta n(x, y)}{n_0} \propto \frac{dI(x, y)}{<I>} \qquad (2)$$

Where $n_0$ and $\Delta n (x, y)$ denote the average refractive index of the fluorescent molecules and its fluctuation at $(x, y)$ position (size $dxdy$ and thickness $dz$), respectively. $<I>$ is the average intensity of the confocal images and $dI(x, y)$ is the fluctuation in the intensity at $(x, y)$ pixel position of the confocal image. From this, an optical lattice is a representation of the spatial refractive index fluctuations of the fluorescent molecules inside the sample [26]. Anderson Tight Binding Model (TBM) is well studied in condensed matter physics for describing the disorder properties of optical systems of any geometry and disorder [27]. If we consider one optical state per lattice site, with inter lattice site hopping restricted to the nearest neighbors only, the tight-binding Hamiltonian can be written as:

$$H = \sum_i \varepsilon_i |i\rangle\langle i| + t \sum_{\langle ij \rangle} (|i\rangle\langle j| + |j\rangle\langle i|) \qquad (3)$$

Where $\varepsilon_i$ is the optical lattice potential corresponding to the $i^{th}$ lattice site, $j^{th}$ is the nearest neighbor of $i^{th}$ lattice site, and $t$ is the inter-lattice site hopping strength. The average IPR value for a lattice system is calculated from the eigenfunction and eigenvalue of above Hamiltonian as

$$\langle IPR \rangle = \frac{1}{N} \sum_{i=1}^{N} \int_0^L \int_0^L E_i^4(x, y)\, dxdy \quad . \qquad (4)$$

Where $E_i$ is the $i^{th}$ eigenfunction of the Hamiltonian of an optical lattice of size $L \times L$, having $N$ lattice point. The disorder in heterogeneous light transparent media such as the biological systems can be specified by two parameters, refractive index fluctuations $\Delta n$ and its spatial fluctuation correlation length $l_c$ and can be expressed as disorder strength, $L_d$:

$$L_d = \langle \Delta n \rangle \times l_c \qquad (5)$$

and the average value of IPR represents the measure of the disorder strength, so IPR and disorder strength can be written as [5,23],

$$\langle IPR \rangle \propto L_d = \langle \Delta n \rangle \times l_c . \qquad (6)$$

### 3. Results

*3.1. Results of partial wave spectroscopy (PWS) of brain tissues*

The engineered finer focusing PWS technique and molecular specific light localization, IPR technique were used to detect the nanoscale structural alterations in the pups' brain tissues and nuclei due to fetal alcoholism. For this, the PWS analysis was performed in the paraffin embedded 10µm thick the control fed

(PF) and ethanol fed (EF) mice pup brain tissue sections to calculate the degree of disorder strength, $L_d$, as mention in section 2.2. These pups' brain tissues/cells were collected at the postnatal day 60 to quantify their structural properties due to fetal alcoholism. Then, using the IPR analysis, the average IPR, $<IPR>$ value was calculated to study the effect of alcohol in EF mice pups brain cells nuclei in the confocal images for all the categories as explained in section 2.2.4. That means the IPR study was performed between (i) control fed (PF) DAPI stained brain cells and ethanol fed (EF) DAPI stained brain cells, and (ii) control fed (PF) H3K27me3 stained brain cells and ethanol fed (EF) H3K27me3 stained brain cells.

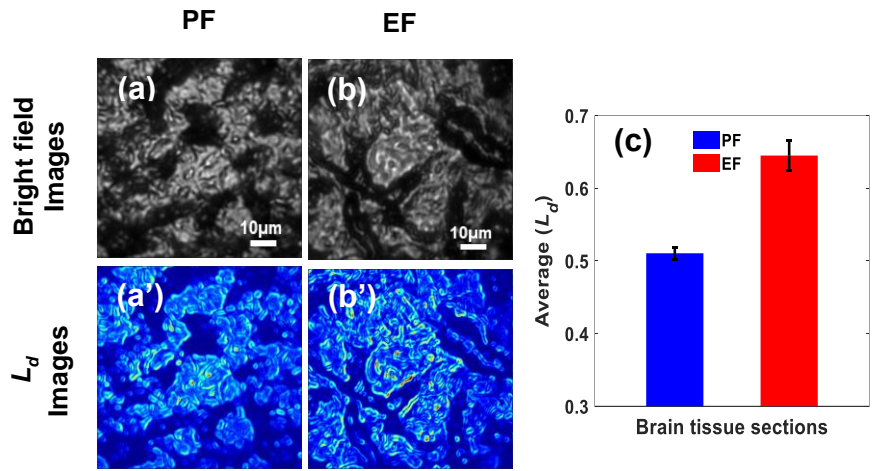

**Fig. 1**: **Brain tissue structural disorder using PWS:** **(a)-(b)** are the bright field images of PF and EF mice pup brain tissues while **(a')-(b')** are their respective $L_d$ images which are distinct than the bright field images. As can be seen from fig **(c)**, the PWS analysis of EF mice pup's micron size brain tissues have a higher degree of the average disorder strength, $L_d$ than PF pups. The average $L_d$ of EF mice pup brain tissue increases by *26%* in reference to the PF pups. (P-value < *0.05*).

Fig.1 represents the bright field images (a)-(b) and $L_d$ images (a')-(b') of PF and EF mice pups thin brain tissue section. Based on quasi 1D approximation, the samples are virtually divided into a number of parallel channels and the backscattered signals propagating along 1D trajectories are collected to calculate the degree of disorder strength at every pixel position. This degree of disorder is represented by a 2D color map in the $L_d$ image where the red color represents the higher disorder strength present in the thin tissue structure averaged along the z-axis. Although the bright field images look similar, the $L_d$ images show that the EF pup brain tissue structure has more red spots indicating a higher degree of disorder strength than the PF. Further, the average disorder strength is calculated and represented in Fig 1(c). The PWS quantification shows that the average of disorder strength, $L_d$, of EF pup brain tissue increases by 26% compare to the PF pup brain tissue. This increase in the degree of disorder strength of EF pup brain tissue structure is due to an adverse effect of alcohol in the brain tissue of pups that carries from the alcoholics' mother in the same amount. Alcohol affects the different components of brain tissues such as chromatin, astrocytes, microglia, protein, lipids, etc. which are initially at the nanoscale level in cells/tissues structural followed by a decrease in the efficacy of brain functioning. Fetal alcohol syndrome and other alcohol related neurodevelopmental disorders in newborn infants are the long-term outcomes of fetal alcoholism. This quantitative approach in the measurement of such structural alterations in brain tissue due to fetal alcohol exposer (FAE) using the PWS technique provides a better understanding of the structural properties at the earliest stages of fetal alcoholism to employ better treatment modalities.

*3.2 Results of Confocal-IPR for: (i) DNA molecular specific DAPI staining and (ii) histone molecular specific H3K27me3 staining:*

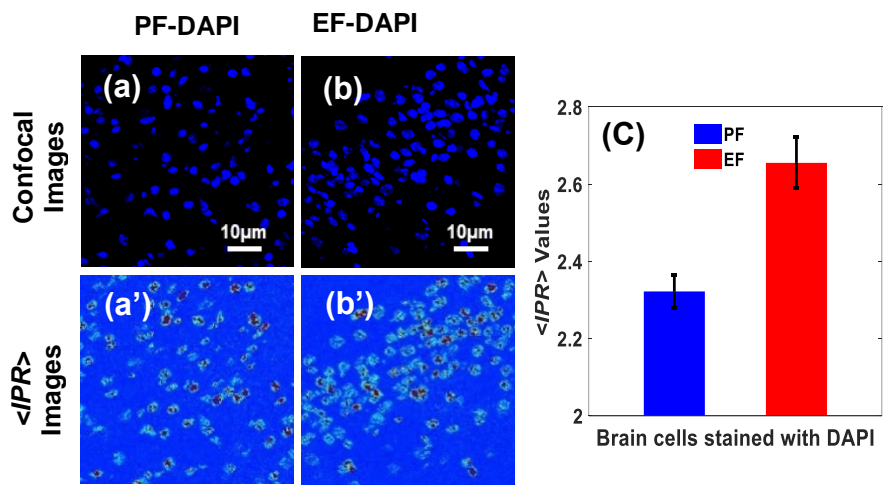

**Fig.2: DNA molecular structural disorder by confocal-IPR: (a)**-**(b)** represent the confocal images of the control (PF) DAPI stained and ethanol fed (EF) DAPI stained brain cells while **(a')**-**(b')** are their respective IPR images. In fig. 2. **(c),** the *<IPR>* values for PF DAPI stained and EF DAPI stained pups' brain cells are presented. The <IPR> value for EF DAPI stained brain cells is found to be higher compared to that of PF DAPI stained brain cells. As DAPI binds with DNA molecules, the above bar plots show that fetal EF pups have more disorder in their brain cell nuclei leading to different kinds of abnormalities. The percentage difference of the *<IPR>* values between these two groups is 14%. (P-value *<0.05, n=10*).

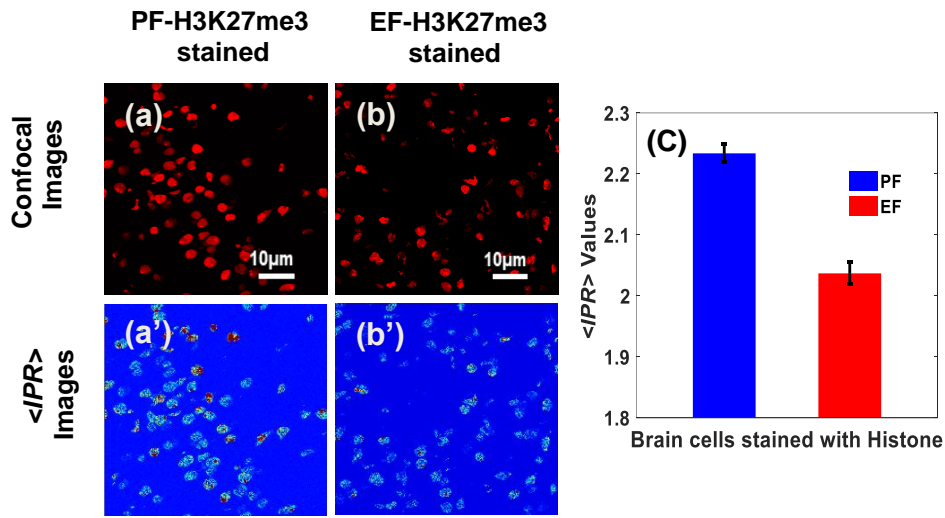

**Fig.3: Histone molecular structural disorder by confocal-IPR:** **(a)**-**(b)** represent the confocal images of the control (PF) H3K27me3 stained and ethanol fed (EF) H3K27me3 stained brain cells targeting histone and **(a')**-**(b')** represent their respective IPR images. In fig.3. **(c)**, the <IPR> values or the degree of structural disorder for PF H3K27me3 stained and EF H3K27me3 stained pups' brain cells are presented. The <IPR> value for PF H3K27me3 stained brain cells is found to be lower compared to that of the PF H3K27me3 stained brain cells. The percentage difference of the <IPR> values for structural disorder between these two groups is 10%. (P-value < *0.05*, *n=10*).

Figs 2. **(a)** and **(b)** show the DAPI stained brain cells for PF and EF respectively, and **(a')** and **(b')** are the corresponding <IPR> or structural disorder strength of these cells. DAPI staining mainly targets DNA molecules in the brain cell nuclei. Fig.2. (c) is the comparison of ensemble <IPR> values of PF and EF DAPI stained brain molecules. The result shows the higher <IPR> value for EF brain cells indicating higher structural disorder than the PF brain cells. This indicates that <IPR> value of DNA increased due to fetal alcoholism. Figs 3. **(a)** and **(b)** show that H3K27me3 stained brain cells for the PF and EF respectively while **(a')** and **(b')** are the corresponding IPR images. H3K27me3 staining mainly targets the histone molecules in the brain cell nuclei. Fig. **(c)** shows the bar graph comparisons of ensemble <IPR> values for PF H3K27me3 stained and EF H3K27me3 stained brain cells. The graphs show histone structures in brain cells exposed to fetal alcoholism has less structural disorder relative to the control. This is opposite to the nuclear DNA molecular structure, which shows the increase in structural disorder in fetal alcoholism.

    The <IPR> values for DNA molecules of EF pup brain cell nuclei are found to be greater than the control. This <IPR> value is correlated with the structural disorder which indicates that the spatial

distribution of DNA molecules in EF pup's brain cells has more structural disorder. The introduction of alcohol results in the spatial variation in the components of DNA and hence, responsible for the changes in the genetics to introduce different fetal alcoholic syndrome.

The <IPR> value of the ethanol fed H3K27me3 stained brain cells is found to be less than the <IPR> value of the control fed H3K27me3 stained brain cells. This suggests that the change in the degree of structural disorder strength of the ethanol fed pub's stained brain cells is less than the control one. Therefore, less mass density fluctuations in the histone proteins of brain cells due to fetal alcoholic exposition conclude suppression in certain gene expression resulting in fetal alcohol syndrome.

### 4. Conclusions and Discussions

The effects of chronic alcoholism during pregnancy, or fetal alcoholism on new born pups' brain are studied using the mesoscopic physics-based dual spectroscopic approach, the PWS and IPR techniques, in a mouse model. First, we quantified the mass density or refractive index fluctuations at the nanoscale level in PF (control) and EF (ethanol fed) mice pups brain tissues and measured the degree of structural disorder strength, $L_d$. The PWS result shows a higher degree of disorder strength of brain tissues from mice pups exposed to fetal alcoholism relative to control pups. Further, the confocal-IPR technique is used to probe the spatial structural disorder properties of molecular mass density of two types of molecules: i) DNA and ii) histone. The results show the degree of molecular structural disorder of DNA increases while that of histone decreases.

As the change in spatial structural disorder is occurring to tissue as well as the molecular mass densities of DNA and histones, indicating the abnormalities in brain functions. This may result in fetal alcohol syndrome and alcohol related neurodevelopmental disorder in infants as long-term effects of fetal alcoholism. The molecular structural changes in DNA and histone may be due to DNA methylation [3]. The change is the overall structural disorder of brain tissue is due to all components of the tissues, including cells and its molecular mass densities.

***Probable reasons for opposite structural changes in DNA and histone molecules in fetal alcoholism:*** Fetal alcohol exposure might have caused the histone protein modifications which in turn could be responsible for enhancing the relaxation of chromatin causing fewer mass density fluctuations and hence a smaller <IPR> value. H3K27me3 is the trimethylation of lysine 27 on histone protein and the addition of alcohol causes the loss of the H3K27me3 methylation process. This loss of methylation may play a role in DNA unwinding and gene expression. This decrease in the <IPR> value or the disorder strength may be due to the loss of methylation from the loosely attached histones but not from the tightly attached histone to the DNA. Furthermore, fetal alcoholism could result in the loose holding of nucleosomes which would cause a more relaxed state of chromatin. The relaxed state of chromatin might increase the homogeneity in histone, reducing the disorder. Because of these changes, upregulation or downregulation in the expression of certain gene types could have resulted in fetal alcohol syndrome disorder in the infant's brain.

In summary, we probed the change in tissue structures, DNA and histone of mice pups, exposed to fetal alcoholism. Preliminary results are promising, and this invites more detailed study to interconnect these results, and also how they relate to brain abnormalities.

**Acknowledgments**

Part of this work is partially supported by National Institutes of Health (NIH) grants (R01EB003682 and R01EB016983) and Mississippi State University to Dr. Pradhan; and AA12307 to Dr. Rao.